# Optical Detuning Strategies for Shielded Loop Resonators

Jakob Gerlach[1*], Reza Aghabagheri[1], Zining Liu[1], Shuai Liu[1], Morteza Teymoori[2], Çağlar Ataman[2], Michael Bock[1], Ali Caglar Özen[1]

[1]*Division of Medical Physics, Department of Radiology, University Medical Center Freiburg, University of Freiburg, Freiburg, Germany*

[2]*Microsystems for Biomedical Imaging Group, Department of Microsystems Engineering, University of Freiburg, Freiburg, Germany*

**2757**/2800 Words

*Correspondence:

Jakob Gerlach

Division of Medical Physics, Department of Radiology, University Medical Center Freiburg, University of Freiburg, Freiburg, Germany.

Email: jakob.gerlach@uniklinik-freiburg.de

# ABSTRACT

**Purpose:** To compare detuning performance and evaluate power requirements of optical detuning methods and to demonstrate feasibility of an optically-detuned four-channel receive array.

**Methods:** Four optical detuning methods were compared in simulations, bench tests, and phantom measurements at 3T against conventional galvanic detuning. Passive detuning was also tested as an additional wireless detuning option. Optical power requirements for the detuning networks were investigated, and a flexible optically-detuned 4-channel shielded-loop resonator (SLR) array was constructed and tested in vivo.

**Results:** A photodiode-PIN diode combination exhibited the highest unloaded $Q$ (68.6) and $Q$ ratio (1.9), with detuning performance and signal-to-noise ratio comparable to that of galvanic detuning at an optical power of 10 mW. Using this detuning strategy, in vivo images of the knee and brain were successfully acquired with a 4-channel flexible array.

**Conclusion:** Optical detuning is a practical alternative to conventional galvanic detuning in flexible SLR arrays. With developments in optical signal and power transmission, optimizing optical detuning while meeting manageable power requirements is an important step toward fully optical receive coil arrays. This study provides a baseline for the total optical power required for active detuning in such optical coil systems.

# INTRODUCTION

In clinical MRI systems, local receive (Rx) coils are connected to the receiver system via coaxial cables. Particularly in dense receive arrays, galvanic cable connections can introduce problems such as cable clutter, signal crosstalk, and interactions with the radio frequency (RF) excitation pulses. Although benefits in SNR and acceleration factor have been demonstrated for a simulated 256-channel array [1], in practice, the maximum number of channels in a head coil has been limited to 128 [2]. To overcome these challenges, optical methods have been proposed [3], [4], [5], [6], [7], [8], [9], which are insensitive to RF fields. However, in most implementations, the optical links were limited to data transfer, while active coil detuning during RF excitation and powering of amplifiers and other active components remained galvanic.

Recently, a fully optical signal and power transmission system was proposed [8], [9]. Here, detuning was also realized optically using a phototransistor, replacing the conventional active detuning. Alternatively, Memis et al. implemented an optically controlled detuning which was powered by on-coil batteries [3]. Wong and co-workers used a photoresistor and a manually switched light source to tune and detune a micro-coil for the localization of interventional devices [10]. Weiss et al. installed a resonant marker with a photodiode at the catheter tip for tracking [11]. Korn et al. employed optical detuning of a surface coil for motion compensation using a self-gating technique [12]. They introduced a configuration in which a photodiode supplies the current to switch the PIN diodes in the detuning circuit. A variation of this configuration was also applied by Saniour et al. for an endoluminal coil [13]. Despite this large variety of optical detuning strategies for single-channel coils, the application of optical detuning to multi-channel arrays has not been investigated. Moreover, previous works implemented different optical detuning methods on diverse coil types and geometries, preventing a direct comparison of detuning performance and image signal-to-noise ratio (SNR). A direct comparison of optical detuning methods using an identical coil element has therefore been lacking.

In this work, we examine different optical detuning strategies on flexible shielded loop resonators (SLR) [14], [15], [16]. SLRs are preferred as the Rx coils, as they are relatively insensitive to the changes in geometry and form, and their resonance behavior is determined mainly by the length of the transmission line (e.g., a coaxial cable or a strip-line) [17], [18], [19], [20], [21], [22], [23], [24], [25]. This property makes SLRs ideal for implementing dense, flexible Rx arrays, and arrays with up to 64 channels have been recently demonstrated [24]. In literature, most commonly double-PIN-diode detuning positioned at the feed port of the SLR has been used [17], [26]. In [21], PIN-diodes at the gap was also tested. We used a single-sided detuning network to identify the configuration that provides optimal detuning and Q-factor. A detailed circuit model for the SLRs combined with the detuning networks was developed and validated through simulations, bench measurements, and phantom experiments at 3T. The minimum optical power was determined to achieve detuning comparable to conventional circuits. Finally, the most effective optical detuning strategy was integrated into a four-channel flexible SLR array and demonstrated in vivo.

# METHODS

The following detuning methods were realized to detune a SLR: a conventional galvanic detuning, passive detuning with crossed diodes, and 4 different optical detuning methods (Fig. 1A). In Table 1, resistance and capacitance of the detuning components are listed. As the main reference, conventional galvanic active detuning was implemented with a PIN diode (MA4P7470F-1072T, MACOM, Lowell, MA). Here, the MR system provides a 10 V DC signal at up to 70 mA to forward bias the PIN diode, creating a short between inner and outer conductor of the SLR and, thereby, a resonance frequency shift. The DC signal path is separated from the RF path using RF choke inductors (1812CS-333XJLC, Coilcraft, Cary, IL, USA) with $L$ = 33 μH and a self-resonance frequency of $f_{SRF}$ = 20 MHz. Passive detuning, which is used either as an additional detuning measure in conventional receive coils or the sole detuning method in inductively coupled coils, was realized with crossed diodes (MADP-007433-0287DT, MACOM) across the gap of the SLR. Here, the RF-induced currents during excitation switch the diodes leading to a resonance shift. Therefore, passive detuning is effective only during RF pulses with a sufficiently large flip angle and depends on the geometry of the receive coils.

In the first optical detuning method a photodiode (PD) was used that generates a photocurrent approximately proportional to the optical power and detunes the coil. Here, the PD (SFH-203-P, OSRAM, Munich, Germany) was integrated directly into the RF path, and the optical control signal was supplied over a fiber eliminating the need for a separate DC path (Fig. 1A). Alternatively, detuning with a phototransistor (PT) was implemented. Upon illumination, the PT conducts between emitter and collector. In this configuration, PTs were placed directly in the RF path, either between the inner and outer conductors of the SLR at the feed port (PT1: SFH-3310, PT2: SFH-309-4, PT3: SFH-309-5/6; OSRAM, Munich, Germany), or across the gap (D-PT: SFH-3310). PTs were also energized optically via fiber without a separate DC control path. Finally, a PD and a PIN diode were combined (PD+PIN) using choke inductors, similar to galvanic detuning, to isolate the PD from the RF path. The bias condition of the PIN diode was different from the galvanic detuning, as the voltage produced by the PD reaches only 0.6 V, even at 50 mW of optical power. At this 0.6 V bias, PD does not fully switch the PIN diode but its resistance drops significantly; therefore, the bias change is sufficient to reach the required RF impedance change for detuning (Supporting Information Tables S1 and S2).

Passive detuning relies on induced current and is therefore most effective across the gap (see Supportive Information Fig. S5C). Active detuning of SLRs is typically achieved by shorting outer and inner conductor [17], [26]. Most of the optical detuning networks adhere to this conventional positioning for active detuning. The one exception is D-PT, which places the same PT as in PT1 across the gap and allows to check the preferential positioning. Unlike [17], [21], [26], we used single-PIN-diode networks for galvanic and PD+PIN detuning. The motivation was to achieve a fair comparison between optical detuning methods. Furthermore, the PT-based detuning was possible using a single fiber connection as in PD+PIN network, instead of two fibers. A comparison of single- vs. double-PIN-diode detuning is summarized in Supporting Information Figure S5A.

To assess the impact of different detuning strategies under controlled conditions and guide the experimental design, RF circuit simulations were performed using ADS (Advanced Design System, 2025, Keysight, Santa Rosa, CA, USA). An SLR was chosen as the receive element and modeled using two coaxial transmission-line (TL) segments, each with an additional inductance across the outer conductor, representing the loop inductance (Figure 2A). The inner conductors of the TL segments were connected, while the outer conductors were left separate, representing the gap. Tuning and matching capacitances in a $\pi$-configuration were assigned as optimization parameters within the nominal range of the physical components (2-10 pF variable capacitors). In practice, the variable matching capacitors are never exactly equal. In our simulations, CM1 and CM2 are therefore independent parameters and a small difference (<0.1 pF) between them results from optimization. The detuning components were modeled as lumped R-C elements in series, connected between the inner and outer conductors of the SLR at the feed port. Circuit models of the detuning elements were based on $R$, $L$, and $C$ measurements performed at 123.2 MHz (measured $^{1}$H Larmor frequency of the MRI system) using an impedance analyzer (4396B, Agilent, Santa Clara, CA, USA). The full ADS schematic, including simulation parameters in the PD+PIN case, is given in Supporting Information Figure S1A.

For bench tests and MRI measurements, an SLR was constructed by forming a loop from a 14.9 cm long semirigid cable (SUCOFORM_47_CU_LSFH, Huber+Suhner, Herisau, Switzerland) and cutting a 2 mm wide gap in the outer conductor opposite to the feed port (Fig. 1B). Variable capacitors (JR100, Knowles Voltronics, IL, USA) were used for tuning and capacitive matching. To compare the detuning methods in the same SLR, networks were connected to the same coil via 0.1-inch pins and sockets. This allowed quick dis- and reconnection of networks without changing coil geometry or position. Loaded and unloaded $Q$ were determined from S11 curves with a vector network analyzer (P5020A, Keysight, Santa Rosa, CA, USA). S21 measurements of tuned and detuned states were obtained using a decoupled pair of sniffer loops at a fixed distance from the coil (see Supportive Information Fig. S7). The inter-probe coupling 3 cm above the phantom was -64 dB at the $^{1}$H frequency (compare [27]).

Based on the best optical detuning method (PD+PIN), a 4-channel flexible SLR array was constructed using a flexible cable (1001935047, TEMP-FLEX, Molex, Lisle, IL, USA). In the array each SLR was a 5.5 cm diameter circular loop (Fig. 1B), soldered directly to the PCB accommodating the detuning network. The PD was fixed in a coupler (Fig. 1C) attached to an optical fiber (MM 50/125 OM2, FS, New Castle, DE, USA). Geometric decoupling by overlapping was applied to decouple nearest-neighbor elements in addition to the intrinsic decoupling of the SLRs, as they operate on their first resonance mode [22]. The coil array was attached to medical-grade synthetic leather, which was clamped on a 3D-printed holder to form a flexible headrest.

All MR measurements were performed at a clinical 3T system (MAGNETOM CimaX, Siemens Healthineers, Forchheim, Germany). The coils were connected to the MRI system via a 4-channel interface circuit with integrated preamplifiers (Flex Coil Interface, Siemens Healthineers). A system diagram of the optically detuned 4-channel system is presented in Figure 1D, photos of the setup on the patient table are shown in Supporting Information Figure S.2. A 850 nm laser (RLTMDL-852-1W-3, Roithner LaserTechnik, Vienna, Austria) was sent to an optical switch (MEMS-1X2-M-85-M2-5-09-05-FP, Fibermart, Hong Kong), which provides an isolation of 35 dB. A trigger signal was programmed in the pulse sequence to activate the switch and deliver optical power to the photodiodes or phototransistors during the RF excitation pulse, detuning the coils. In the array measurements a 1:4 splitter (FM-MM-104ABSS, Fibermart, Hong Kong) distributed the light equally to each channel.

An optical power meter (PM101 with S145C power head, Thorlabs, Newton, NJ, USA) was used to monitor the laser power. For the bench tests the optical power was measured at the site of detuning, in the MR measurements it was measured at one switch output. Fibers and connectors between switch and site of detuning accumulate optical losses of up to 20%. Optical powers in the analysis are presented as measured at the switch output, not the reduced power arriving at the site of detuning.

For single-coil characterization, a bottle phantom (3.75 g $NiSO_4 \cdot 6H_2O$, 5 g NaCl per 1000 g $H_2O$, height = 24.5 cm, diameter = 12 cm) was measured using a 2D-FLASH sequence (TR/TE=100/10ms, α=25°, FOV=250×250mm$^2$, ΔV=1×1×5mm$^3$, $N_{averages}$=5, coil-phantom distance: 3mm). To visualize the detuning, a measurement series was acquired using the body coil in Rx mode while keeping the Rx coils on the phantom. In this configuration residual coupling between the SLRs and the body coil as a result of sub-optimal detuning is expected to cause $B_1^+$-field amplification in the close vicinity of the Rx coils. Flip angle maps were also measured with a single SLR element placed on the phantom for different optical powers. Here, double angle method was used (TR/TE=1500/10ms, α=20°/40°, FOV=200×200mm$^2$, ΔV=1.8×1.8×3mm$^3$, coil-phantom distance: 3mm) with the body coil in Tx/Rx mode. A 3×3 pixel mean filter was applied for visualization. Additionally, to evaluate potential SNR loss due to the detuning network components, image SNR for different detuning networks was compared, using the SLR as receive coil.

In vivo measurements of the knee (2D-FLASH, TR/TE=200/10ms, α=25°, FOV=200×200mm$^2$, ΔV=0.78×0.78×2mm$^3$) and the brain (3D-FLASH, TR/TE=30/8ms, α=25°, FOV=200×200mm$^2$, $N_{slices}$=128,

$\Delta V=0.8\times0.8\times0.8mm^3$, Acc-factor=4) of a healthy volunteer were performed with the optically detuned flexible SLR array. All measurements were carried out in accordance with relevant guidelines and regulations; healthy volunteer imaging was approved by the institutional review board (Ethikkommission) of the University Medical Center Freiburg (No. 160/2000); and informed written consent was obtained before imaging.

## RESULTS

*Q* values from measured and simulated S11 curves (Supporting Information Fig. S1B) have a mean deviation of 12.6% and a maximum deviation of 29.8% for PD (Table 1), which can be attributed to uncertainties in impedance measurements and the parasitic effects that are not included in simulations. An SLR without a detuning network had an unloaded *Q* of $Q_{U,meas}$ = 83.7, which is considered as the reference Q value. Detuning networks introduce additional losses. Among the optical detuning circuits, the PD+PIN network preserved the unloaded *Q* most effectively, with a $Q_{U,meas}$ reduction of 18% compared to the reference. The galvanic detuning network reduced $Q_{U,meas}$ by 10%, other optical detuning networks reduced $Q_{U,meas}$ by 50-80% (Table 1). When the phototransistor is placed across the gap (D-PT), a higher reduction in $Q_{U,meas}$ was measured than between the inner and outer conductor (PT1).

Bench measurements comparing single-PIN-diode and double-PIN-diode detuning networks are summarized in Supporting Information Figure S5A. The double-PIN-diode detuning provided 10 dB higher isolation, while exhibiting a lower loaded/unloaded Q of 23.9/66.6 compared to single-PIN-diode with 36.7/75.3.

The design of the 4-channel array elements was based on the PD+PIN method. In contrast to the PD+PIN network with pin-connection listed in Table 1, the loop from flexible cable was directly soldered to the PCB to create the array elements. They displayed $Q_{U,meas}$ = 74±2 and *Q* ratios of 3.5±0.1. S11 values agreed well with the simulations, considering a 20% uncertainty in the component parameters (Fig. 2B).

In Figure 2C, S21 curves in tuned and detuned states are shown for a subset of detuning circuits. In the galvanic detuning network, the resonance shifts by $\Delta f$ = 23 MHz. For passive detuning with crossed diodes, the S21 curve broadens with increasing power as expected. A similar behavior is observed for the D-PT network, with increasing optical power. In the other optical methods, the resonance shifts to a lower frequency and the peak broadens, i.e. *Q* is reduced. Isolation by detuning, i.e. $\Delta_{S21}$, reached $\Delta_{S21,Galvanic}$ = 22.19 dB for galvanic detuning. With the PD+PIN network, $\Delta_{S21,PD+PIN}$ was 21.31 dB at 20 mW optical power. All optical networks yielded higher isolation with increasing optical power; the isolation of PT and D-PT networks was inferior to PD+PIN network, e.g. $\Delta_{S21,PT3}$ = 14.42 dB and $\Delta_{S21,D\text{-}PT}$ = 10.44 dB at 20 mW. Measured at 123.2 MHz and $P_{opt}$ = 20 mW, the PD+PIN network displayed a resistance of 42 Ω, significantly lower than the PT networks ($R_{PT}$ = 87±18 Ω).

The phantom images acquired with the body-coil and the SLR at phantom surface compares the different decoupling levels (Figure 3A). Without detuning the SLR couples to the body coil producing strong image perturbations, whereas both galvanic and PD+PIN detuning reduced these artifacts. Without detuning, the signal in a circular ROI 1 cm below the phantom surface saturates due to the strong coupling (red circle in Fig. 3A). With increasing optical power ($P_{opt}$ = 0.5/2/5/10/20 mW), detuning improved and relative signal intensities reduced to $SI_{PD+PIN}$ = (273±16)%/(127±3)%/(102±2)%/(97±3)%/(95±3)% of the galvanic reference. In Figure 3B, phantom images were acquired with the SLR: At $P_{opt}$ = 0.5 mW a local signal increase can be observed which is caused by $B_1^+$ field distortions. The relative SNR in a circular ROI 1 cm below the surface for optical powers of $P_{opt}$ = 0.5/2/5/10/20 mW was measured as $SNR_{PD+PIN}$ = (165±17)%/(106±8)%/(105±8)%/(104±8)%/(105±8)% of the galvanic reference. The flip angle maps (Figure 3C) display relative deviations from the nominal flip angle of (+40±23)%/(+42±25)%/(+14±9)%/(-5±9)%/(-2±14)% (mean ± standard deviation inside ROI) at $P_{opt}$ = 0.5/2/5/10/20 mW. Flip angle maps with PD+PIN at $P_{opt} \geq 10$ mW are comparable to the ones with galvanic detuning and the reference without surface coil, which exhibit relative flip angle deviations of (-2±12)% and (+1±12)%, respectively.

In the optically detuned 4-channel array, the strongest coupling among the array elements occurred between the non-adjacent coils 1 and 4 (S14 = 12.64 dB). For all other coil pairs, isolation exceeded 15 dB. Individual elements could be activated separately as confirmed by single-coil measurements with all other elements optically detuned. No signal leakage was detected from deactivated elements (Supporting Information Fig. S3). In vivo images of the knee (Fig. 4A) and the occipital lobe (Fig. 4B-D) show uniform image quality when the coils are optically detuned ($P_{opt}$ = 10 mW per channel) without visible artifacts.

# DISCUSSION

In this work, four different optical detuning methods were implemented and compared with conventional detuning methods. The best optical detuning system, the PD+PIN network, was successfully integrated into a 4-channel SLR array. Although this work demonstrates detuning only with SLRs, phantom measurements on a loop coil [28] confirmed that optical detuning methods can also be applied to other receive coil structures (Supporting Information Fig. S4). A number of modifications to the coil design were investigated, necessitating setup flexibility and the ability to adjust tuning and matching. To build an array with high channel number, long-term reliability will rise in importance, i. e.

strain relief for optical fibers should be implemented and fixed capacitors should be preferred over variable ones.

The PD+PIN network reduced *Q* less than the PD, PT and D-PT networks, which might be attributed to higher capacitance of photoactive components (3.76 – 6.38 pF vs 0.87 pF, see Table 1). In the PD+PIN network, the PD was separated from the RF path via large inductors, isolating its capacitance from the coil. The PD+PIN network achieved a comparable $\Delta_{S21}$ as the galvanic detuning, exceeding 20 dB. According to [29], a detuning efficiency $\Delta_{S21}$ of 20log(Q)+13 dB is required. For this coil setup with a *Q* of 74±2, this would correspond to $\Delta_{S21}$ = 50 dB. However, a $\Delta_{S21}$ = 22 dB was measured for galvanic and $\Delta_{S21}$ = 21 dB for optical detuning, which is significantly lower. The proposed threshold might be conservative, and even commercial coils on the 3T system did not meet it: For example, vendor-supplied 7-cm- and 11-cm-diameter single-loop coils would require $\Delta_{S21}$ = 47 dB and 44 dB, but measured $\Delta_{S21}$ = 27 dB and 20 dB at the bench, respectively. See [30] to compare with $\Delta_{S21}$ achievable with preamplifier decoupling. The lower detuning efficiency values observed in our study may partly result from differences in coil design and experimental setup. The coils used in [29] differ from those in our study in several aspects, including type, geometry, and the distance between the coil and the phantom. To ensure a fair comparison under realistic conditions, we based our evaluation on a galvanically detuned reference using the identical coil and phantom setup.

In the phantom measurements, the PD+PIN network had a detuning performance that was comparable to galvanic detuning even at optical power levels as low as 10 mW. In contrast to170 mW (measured at the MRI system interface during detuning as 100 mA at 1.7 V across the PIN-diode) delivered by the MRI system to the PIN diode in galvanic detuning, the PD+PIN network constitutes a low power alternative. Any extrapolation to large arrays, however, must also consider system level constraints. Bringing 256 fibers into the bore and distributing the optical power introduces potential bottlenecks related to cable routing, connector count, bend radius constraints, and thermal load near a central light source. A practical implementation would likely require an optical power distribution hub, in which one or a few high-power sources feed an internal splitter tree, and each hub supplies a subset of coils (for example, coil groups or modules rather than all 256 elements from a single splitter). Under such an architecture, and assuming that each channel requires on the order of 10 mW at the PD with total losses not exceeding those of the present prototype, the total source power for a 256-channel array would remain on the order of a few watts. We therefore anticipate 3 W total optical power as an order of magnitude feasibility estimate.

During data acquisition, the MRI system applies a -30 V reverse bias to maintain a stable and well-defined tuning state of the PIN diode. In the absence of a galvanic connection to the system's bias circuitry, reproducing such bias levels in optical detuning networks is not straightforward. However, this large negative bias is a conservative engineering choice rather than a physical requirement. A -5 V bias has been applied [31] and a large negative bias is expected to decrease reverse-bias switching time [32]. In our setup, we were limited by switching times of 0.4 ms for tuning and 2 ms for detuning caused by the MEMS-based optical switch. With a faster switch (NanoSpeed 1X2, Agiltron, Woburn, MA, USA) we measured a tuning time of 19.3 µs and a detuning time of 1.7 µs for a SLR with PD+PIN detuning network (see Supporting Information Fig. S8). These values align in their orders of magnitude with the results of Saniour et al. [13], who reported 13.6 µs for tuning and 1.7 µs for detuning of a photodiode and PIN diode combination. Such delays can be considered tolerable for most UTE/ZTE implementations where over 50 µs delays are used [33]. Within this work, we chose to use the slower MEMS-switch for its high isolation (35 dB), as small power leakages to the detuning network can influence the SNR of an MRI measurement significantly.

Switching was controlled by a custom trigger waveform programmed in the sequence. As the source files are not available for many clinical sequences, only a FLASH sequence was used in this work. However, PD+PIN detuning was also tested at α up to 90° (Supporting Information Fig. S6), confirming its functionality for a wider range of sequences including high RF amplitudes.

## CONCLUSION

A photodiode-PIN diode combination (PD+PIN) provides good optical detuning with minimal reduction in *Q* and detuning performance comparable to conventional methods. It is a low-power alternative for detuning Rx coil arrays and can potentially be used in dense Rx arrays. Optical PD+PIN detuning could be combined with optical signal and power transmission techniques to enable fully optical MR receive systems.

## ACKNOWLEDGMENTS

This study was funded by Deutsche Forschungsgemeinschaft (DFG, German Research Foundation) under grant number #1050092401.

# FIGURES

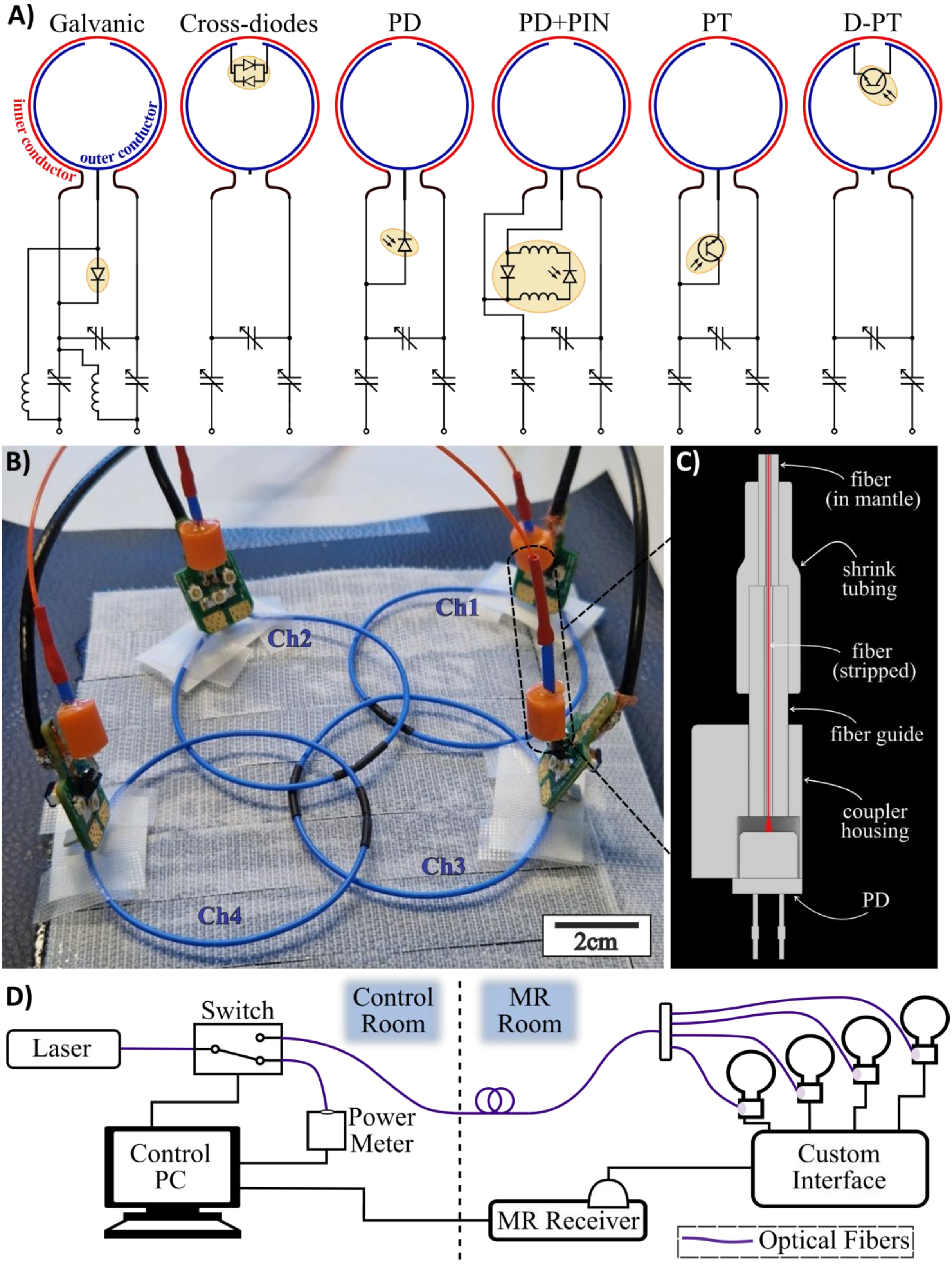

Fig. 1: Schematics and setup. A) sketches of the reference and optical detuning methods examined in this work, highlighting the components essential to the detuning process. B) photo of the four array elements overlapping to achieve geometric decoupling. Detuning signal is sent through optical fibers

coupled to a photodiode which is connected to the tune and match network, C) displays a detailed sketch of the coupler. D) system diagram of an optically detuned receive array MR measurement.

Table 1: Overview of tune and match circuits compared in this work. For each of the circuits, resistance and capacitance at 123.2 MHz are listed for the central detuning component (as highlighted in Fig. 1A) and the *Q* factors as determined by simulations and in bench tests are shown.

| Detuning circuit | Component | | Simulation $Q_U$ | Measurement | | |
|---|---|---|---|---|---|---|
| | *R* [Ω] | *C* [pF] | | $Q_U$ | $Q_L$ | $Q_{Ratio}$ |
| Galvanic | 741 | 0.59 | 72.9 | 75.3 | 36.7 | 2.05 |
| Cross-diodes | 35 | 1.04 | 95.1 | 74.0 | 40.2 | 1.84 |
| PD | 343 | 4.25 | 17.2 | 24.5 | 15.4 | 1.59 |
| PD+PIN | 512 | 0.87 | 62.8 | 68.6 | 36.1 | 1.90 |
| PT1 | 73 | 3.76 | 43.3 | 39.1 | 24.4 | 1.60 |
| PT2 | 72 | 6.38 | 25.4 | 27.5 | 18.1 | 1.52 |
| PT3 | 82 | 6.20 | 24.0 | 26.1 | 19.6 | 1.33 |
| D-PT | 73 | 3.76 | 17.7 | 18.5 | 15.5 | 1.19 |

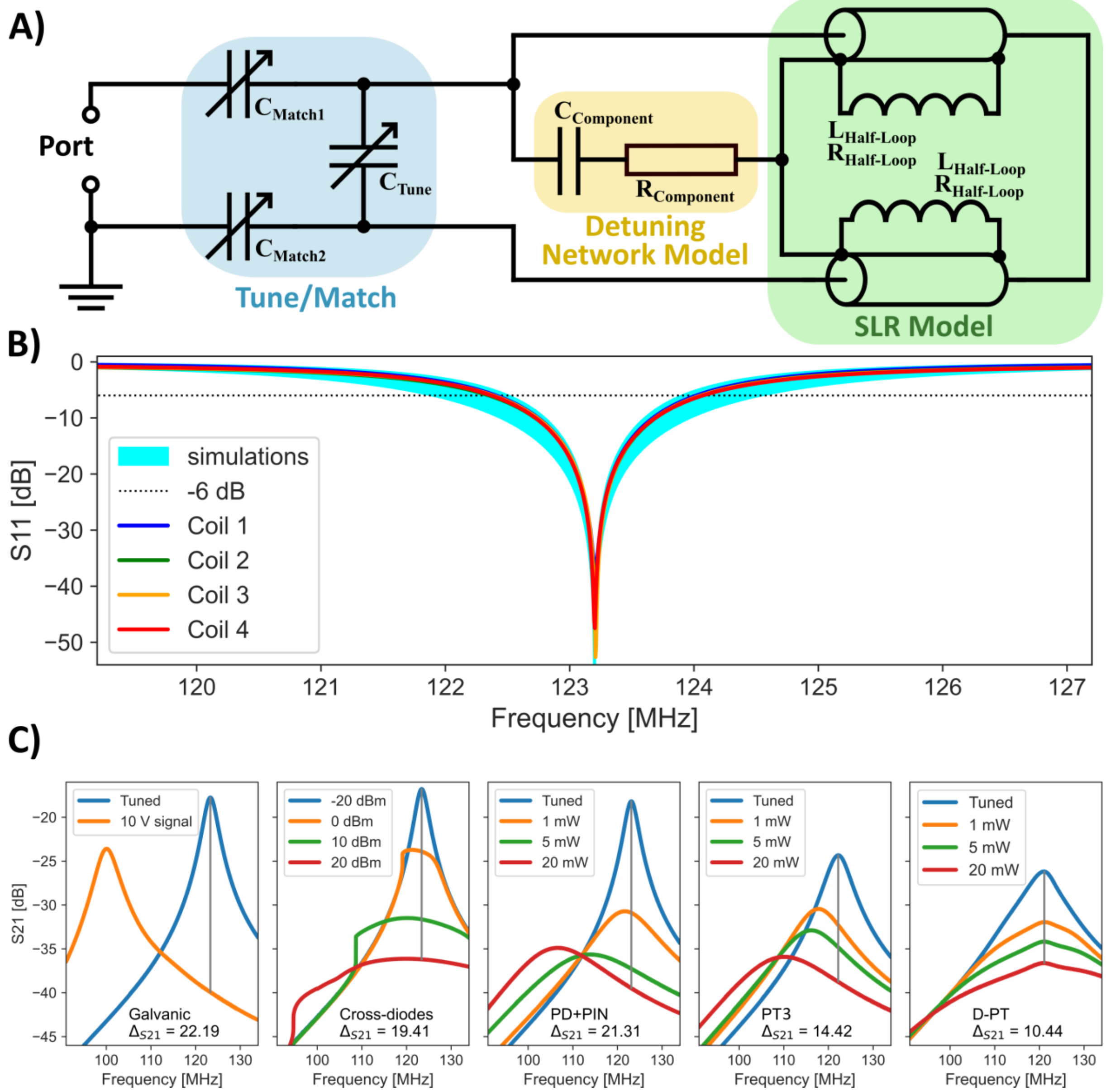

Fig. 2: Simulation and bench test characterization. A) simplified model used for S11 simulations. B) unloaded S11 parameters of the four coils used in the array (curves overlap) in comparison with simulations in a range of loop parameters (R, L) and component parameters (R, C) varying by 20%, respectively. C) measured isolation between tuned and detuned state for 4 selected detuning circuits. S21 curves were measured by connecting a decoupled pair of sniffer loops to ports 1 and 2 of a network analyzer and placing them at a fixed distance from the coil. Each coil was matched to 50 Ω and connected to a Tee-bias circuit to deliver galvanic detuning signal. The relevant parameter distinguishing the curves depends on the type of detuning: electric signal for galvanic detuning, optic illumination power for PD+PIN and PT circuits and VNA output power for passive detuning.

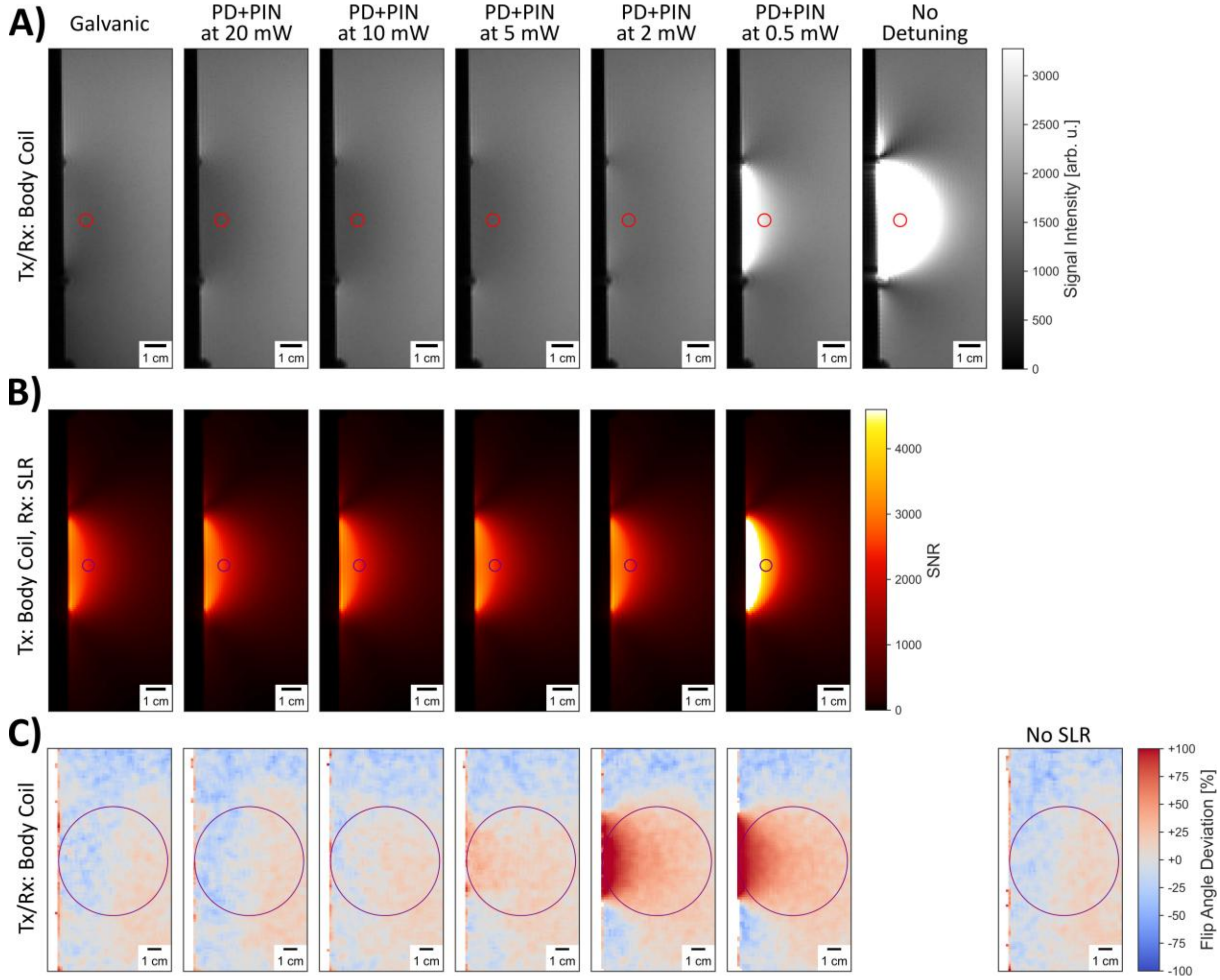

Fig. 3: Bottle phantom measurements. A) Body coil images with surface coil present, shown is signal intensity. For PD+PIN at $P_{\text{opt}}$ = 0.5/2/5/10/20 mW, the mean signal intensity within the ROI is $SI_{PD+PIN}$ = 273%/127%/102%/97%/95% of the galvanic reference. B) SNR maps recorded by the surface coil. For PD+PIN at $P_{\text{opt}}$ = 0.5/2/5/10/20 mW, the mean SNR is $SNR_{PD+PIN}$ = 165%/106%/105%/104%/105% of the galvanic reference (SNR is determined as ratio of pixel value by standard deviation of a noise region outside the phantom). C) Flip angle maps recorded with the double angle method, represented as relative deviation of measured flip angle from nominal flip angle. Optical detuning (PD+PIN) at different illuminations is contrasted with galvanic detuning. For additional references, A) includes a fully tuned coil placed on the phantom (“No Detuning”), C) includes a phantom only measurement without surface coil (“No SLR”).

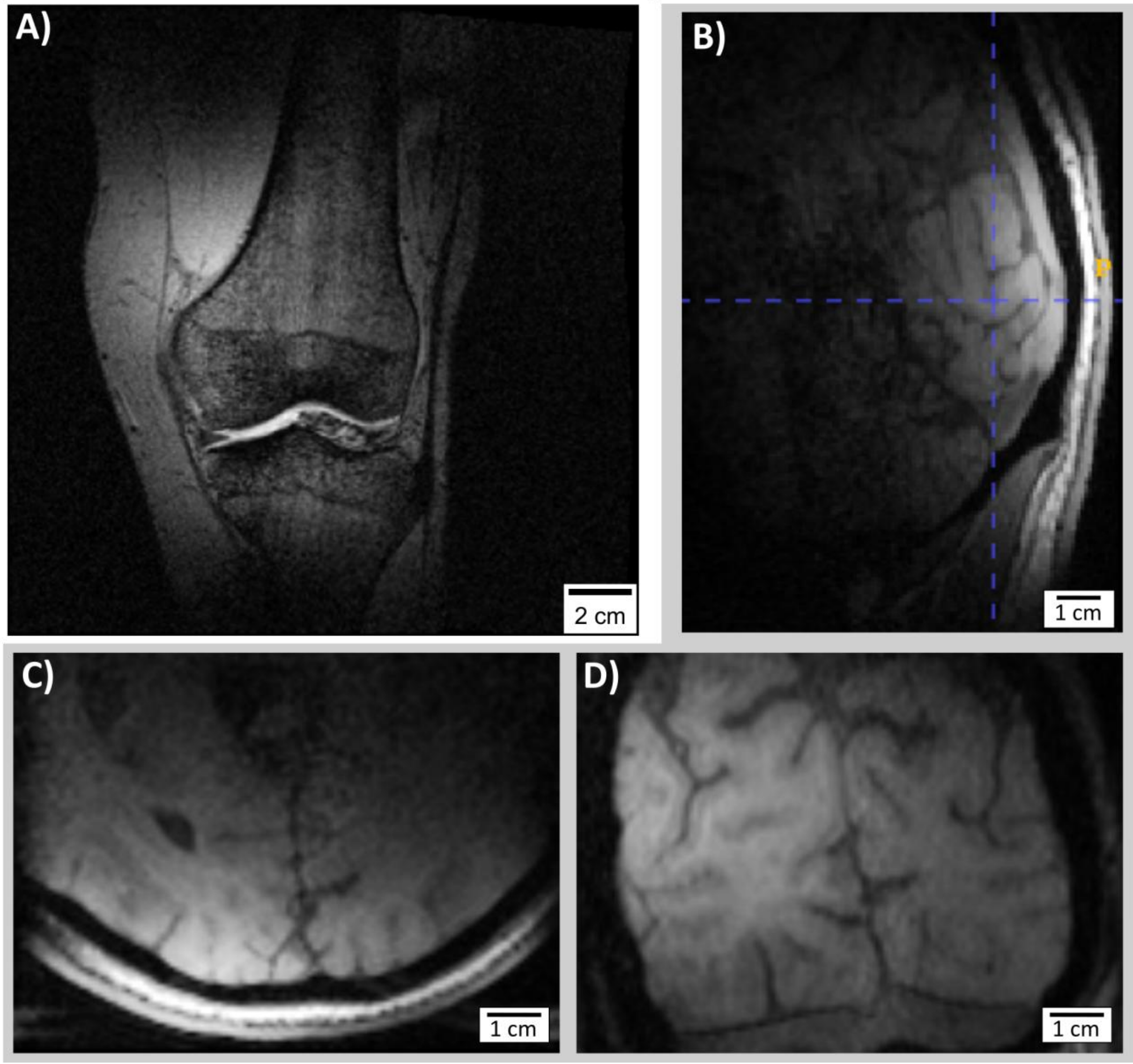

Fig. 4: In vivo measurements recorded with an optically detuned receive coil array. A) Coronal slice of the knee. B-D) Sagittal, transverse and coronal view of the occipital lobe in a 3D data set. In B), "P" marks the posterior direction, the dashed lines indicate image planes of C) and D).

## SUPPORTING INFORMATION

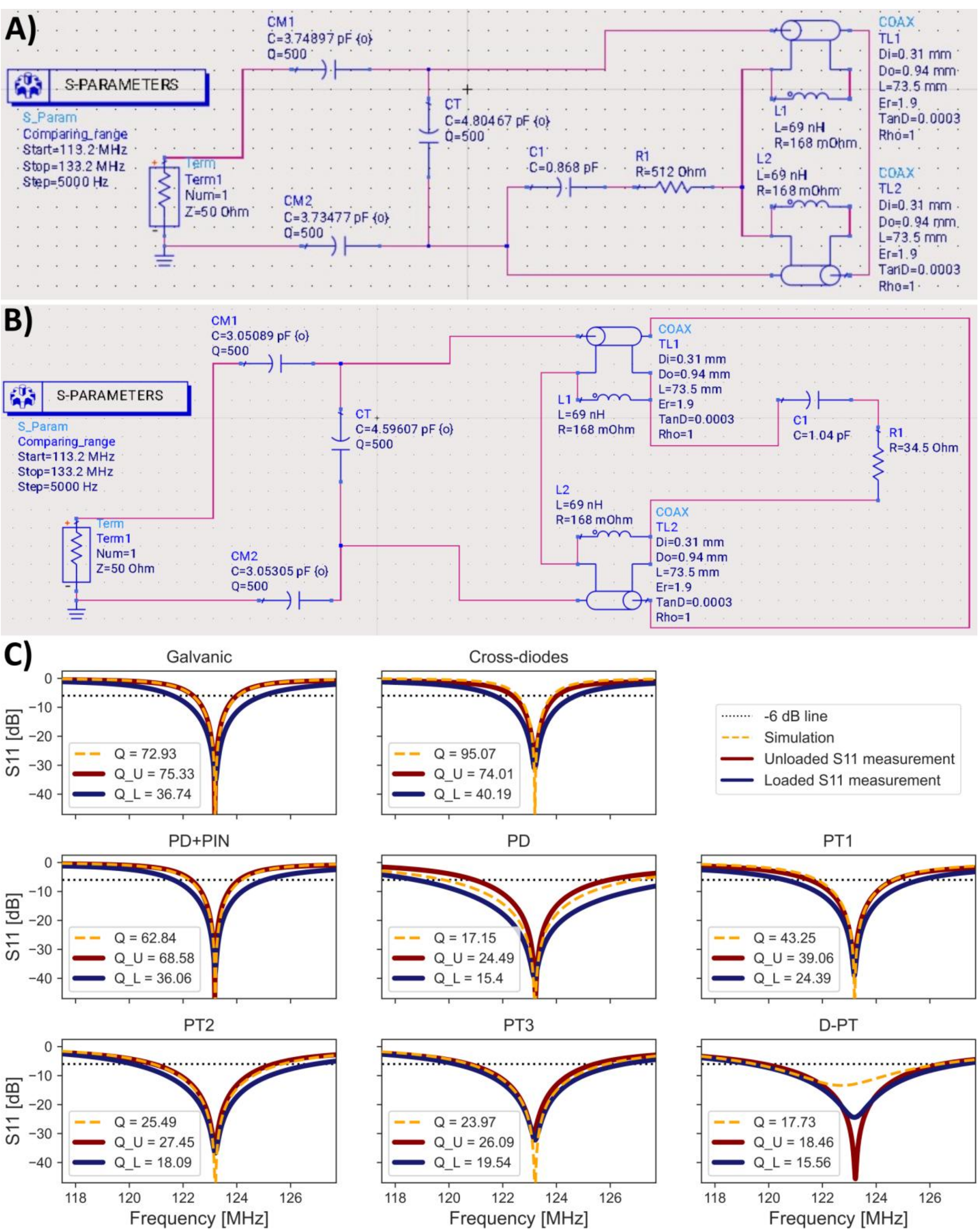

Supporting Information Figure S1: (A) A detailed ADS simulation model used for the simulation of S11 curves (in this specific case of the PD+PIN network). The models vary depending on loop parameters, detuning method and components. Loop *L* and *R* were measured with an impedance analyzer, the transmission line characteristics taken from the datasheet (SUCOFORM_47_CU_LSFH). (B) displays the model in cases, where the detuning network is located across the gap (Cross-diodes or D-PT, this example: Cross-diodes). (C) Composition of measured unloaded and loaded S11 curves for each detuning circuit, contrasted with simulation results.

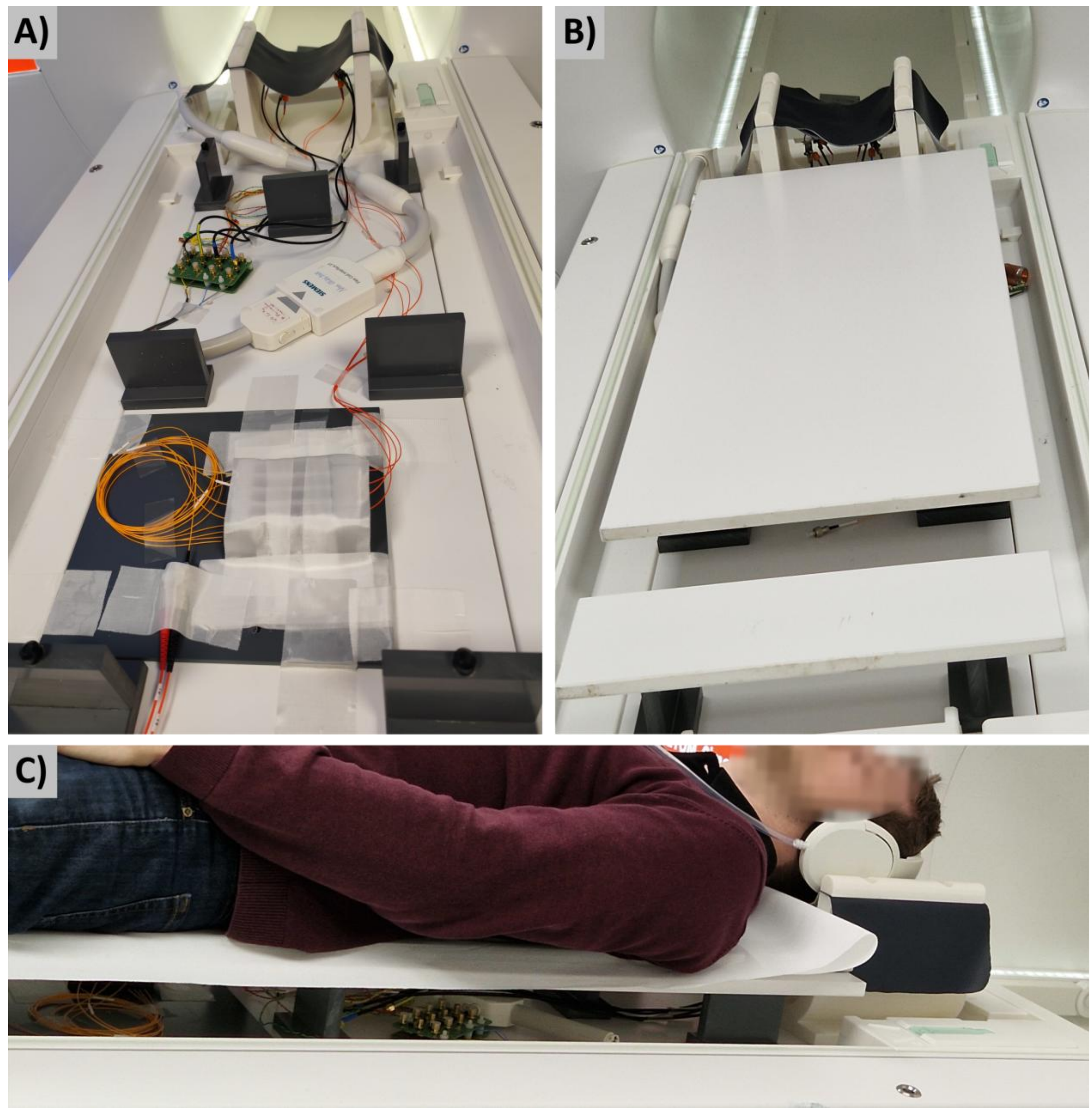

Supporting Information Figure S2: Photos of measurement setup on patient table. A) Open setup displaying interfaces, connections and fibers leading to the headrest. B) Setup covered underneath patient board. C) Side view of volunteer lying on the board, parts of the setup still visible.

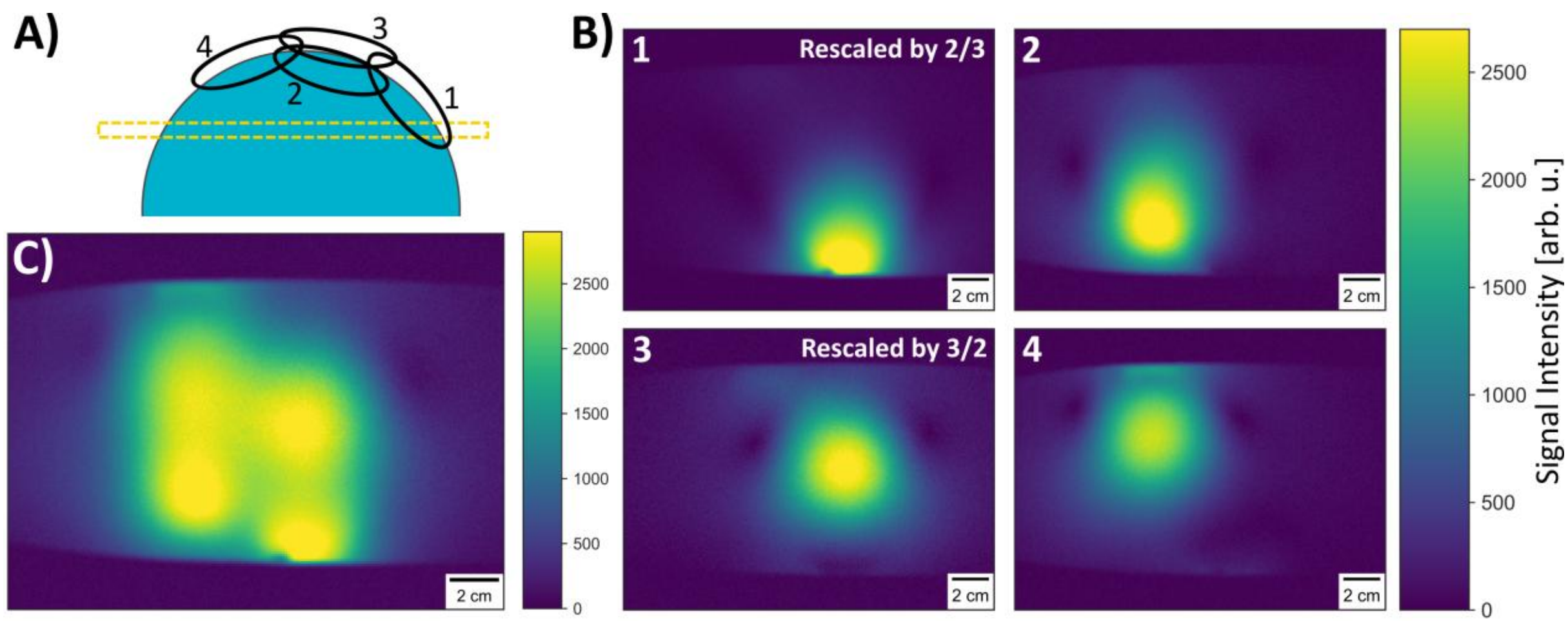

Supporting Information Figure S3: Comparison of individual channels and combined image (2D-FLASH, TR/TE=100/10ms, α=25°, FOV=250×250mm$^2$, ΔV=1×1×3mm$^3$, $N_{\text{averages}}$=5). (A) indicates the placement of the individual coils on the phantom with respect to the measurement plane. (B) presents images acquired by one channel each while the others are detuned, channel 1 and 3 are rescaled to account for the difference in proximity to the measurement plane. (C) displays a phantom measurement with 4 coils (adaptive combination).

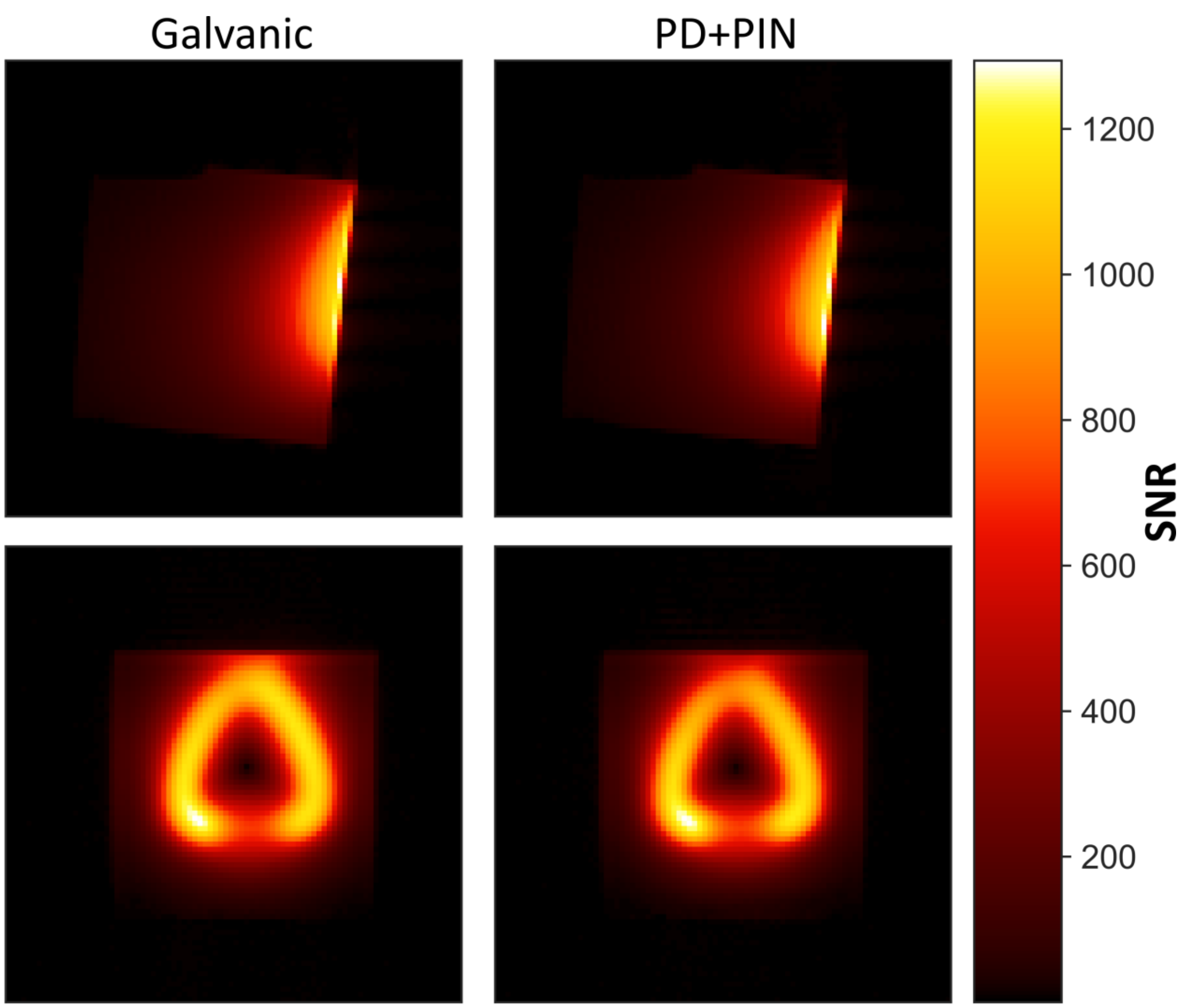

Supporting Information Figure S4: Comparison of SNR of galvanic and optical detuning with the PD+PIN method (2D-FLASH, TR/TE=100/10ms, α=25°, FOV=250×250mm$^2$, ΔV=2×2×5mm$^3$). Galvanic detuning is shown on the left side, PD+PIN detuning on the right side. The top row presents transverse, the bottom row sagittal imaging planes recorded with a loop coil introduced in [28].

## PD output biasing the PIN-diode

In the PD+PIN design presented in this work detuning does not correspond to full activation ("switching on") of the PIN-diode. Instead, the photocurrent and voltage of the PD under irradiation changes the bias of the PIN-diode, altering its resistance to the AC measurement signal. Supporting Information Tables 1 and 2 display bench test results to illustrate the process.

Supporting Information Table 1: $V_{OC}$ and $I_{SC}$ of the OSRAM SFH 203 P photodiode at different optical powers $P_{Opt}$ (852 nm).

| $\boldsymbol{P_{Opt}}$ [mW] | 0.0 | 1.5 | 2.5 | 5.0 | 10 | 20 | 50 |
|---|---|---|---|---|---|---|---|
| $\boldsymbol{V_{OC}}$ [V] | 0.27 | 0.48 | 0.50 | 0.52 | 0.55 | 0.57 | 0.60 |
| $\boldsymbol{I_{SC}}$ [mA] | 0.00 | 0.35 | 0.57 | 1.32 | 3.15 | 7.48 | 17.5 |

For the PIN-diode (MACOM MA4P7470F-1072T), the datasheet lists a maximum forward voltage of $V_F$ = 1.0 V when measured with a forward current of $I_F$ = 100 mA. In a real diode curve, however, the diode is not suddenly switched at this maximum forward voltage, but gradually switched or activated. With a DC source delivering 3 mA input current to the PIN-diode (similar to the PD's $I_{SC}$ at $P_{Opt}$ = 10 mW) the voltage is varied. At 0.739 V, 2.7 mA (90% of the input current) passes the PIN-diode, measured with a multimeter. While the activation or switching of the diode is therefore not a sudden shift, but a gradual change, the voltage supplied by the PD is significantly lower (see Supporting Information Table 1). The resulting resistance of the PD+PIN setup for the AC measurement signal is displayed in Supporting Information Table 2.

Supporting Information Table 2: Series resistance $R_S$ of the PD+PIN setup (including the $L$=33 µH choke inductances separating PD and PIN-diode), measured for varying optical powers $P_{Opt}$ with an impedance analyzer at a frequency of 123.2 MHz.

| $\boldsymbol{P_{Opt}}$ [mW] | 0.0 | 1.0 | 2.0 | 5.0 | 10 | 20 | 50 |
|---|---|---|---|---|---|---|---|
| $\boldsymbol{R_S}$ (@ 123.2 MHz) [Ω] | 512 | 316 | 188 | 98.5 | 60.1 | 41.5 | 27.3 |

A previous work by Saniour et al. (2017) [13] used two PDs in series, which was probably needed to reach the voltage to fully switch the PIN-diode. In our case, the photocurrent of the PD is changing the bias condition of the PIN-diode which was sufficient for detuning.

To isolate the measurement signal (AC) from the detuning signal (DC), the PD+PIN design features $L$ = 33 µH choke inductors (Coilcraft 1812CS-333). Inspite of their higher resistance of $R$ = 13.5±0.1 Ω they are preferable over lower inductance values, e. g. $L$ = 1.2 µH inductors (Coilcraft 1812CS-122) which feature only $R$ = 1.05±0.01 Ω. When replacing the $L$ = 33 µH inductors with $L$ = 1.2 µH inductors, bench tests showed no notable difference (≈1%) in isolation between tuned and detuned state ($\Delta_{S21}$), yet the unloaded $Q$ deteriorates by about 25 % (see Fig. S5B). This shows that high isolation between AC and DC current paths is vital to preserve $Q$.

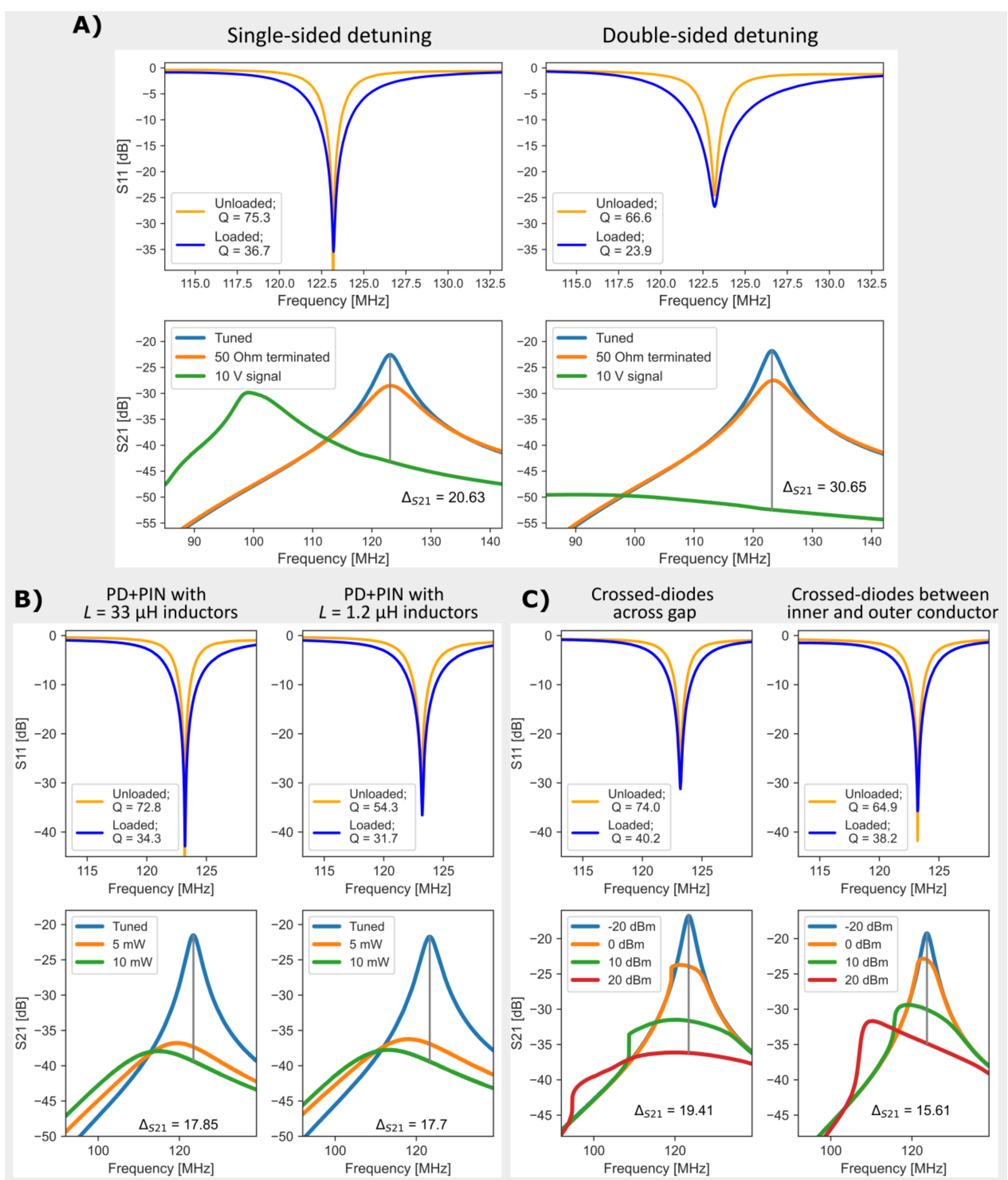

Supporting Information Figure S5: S-parameter comparison for the SLR with various detuning networks. A) Single- vs. double-PIN-diode detuning implemented on the same SLR. It also displays a 6 dB drop in peak S21 when a coil is terminated with 50 Ω. B) PD+PIN detuning network with RF choke inductors of $L$ = 33 µH (left column) and $L$ = 1.2 µH (right column). No significant difference in $\Delta S_{21}$ was observed, while the network with lower inductor resulted in a lower Q of the SLR. C) Performance comparison of passive detuning network using crossed-diodes positioned either across the gap or between inner and outer conductor.

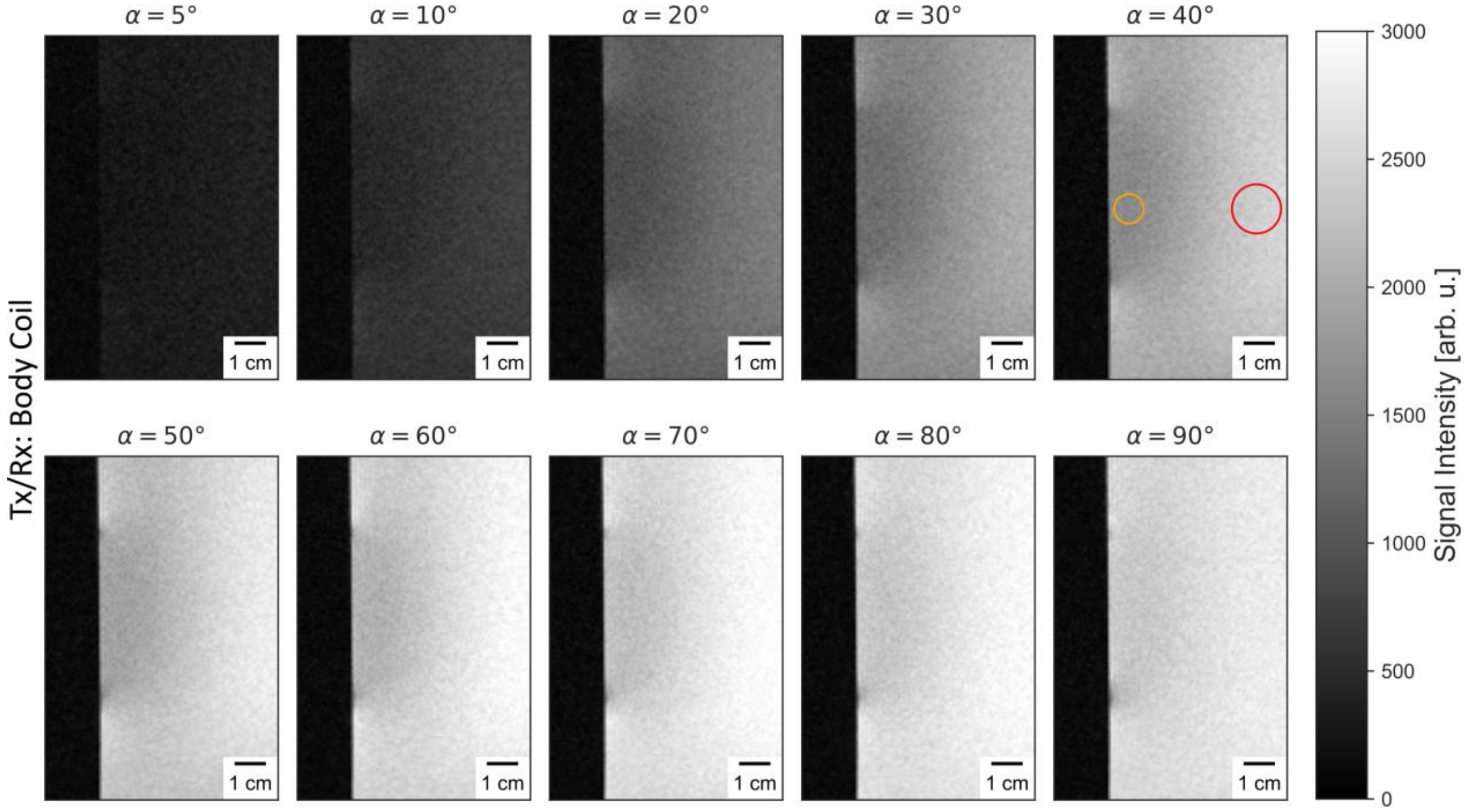

Supporting Information Figure S6: Body coil images with surface coil present, detuned with the PD+PIN method at $P_{opt}$ = 10 mW (2D-FLASH, TR/TE=100/10ms, FOV=200×200mm$^2$, ΔV=0.8×0.8×3mm$^3$). Displayed is signal intensity at several flip angles, illustrating detuning at varying RF pulse amplitudes. Depending on flip angle, the mean signal of a circular ROI close to the detuned surface coil (orange circle in α = 40° measurement) is 61.4% – 85.6% that of an ROI in the center of the phantom (red circle in α = 40° measurement), as compared to 79.0% in a reference measurement without surface coil (not depicted).

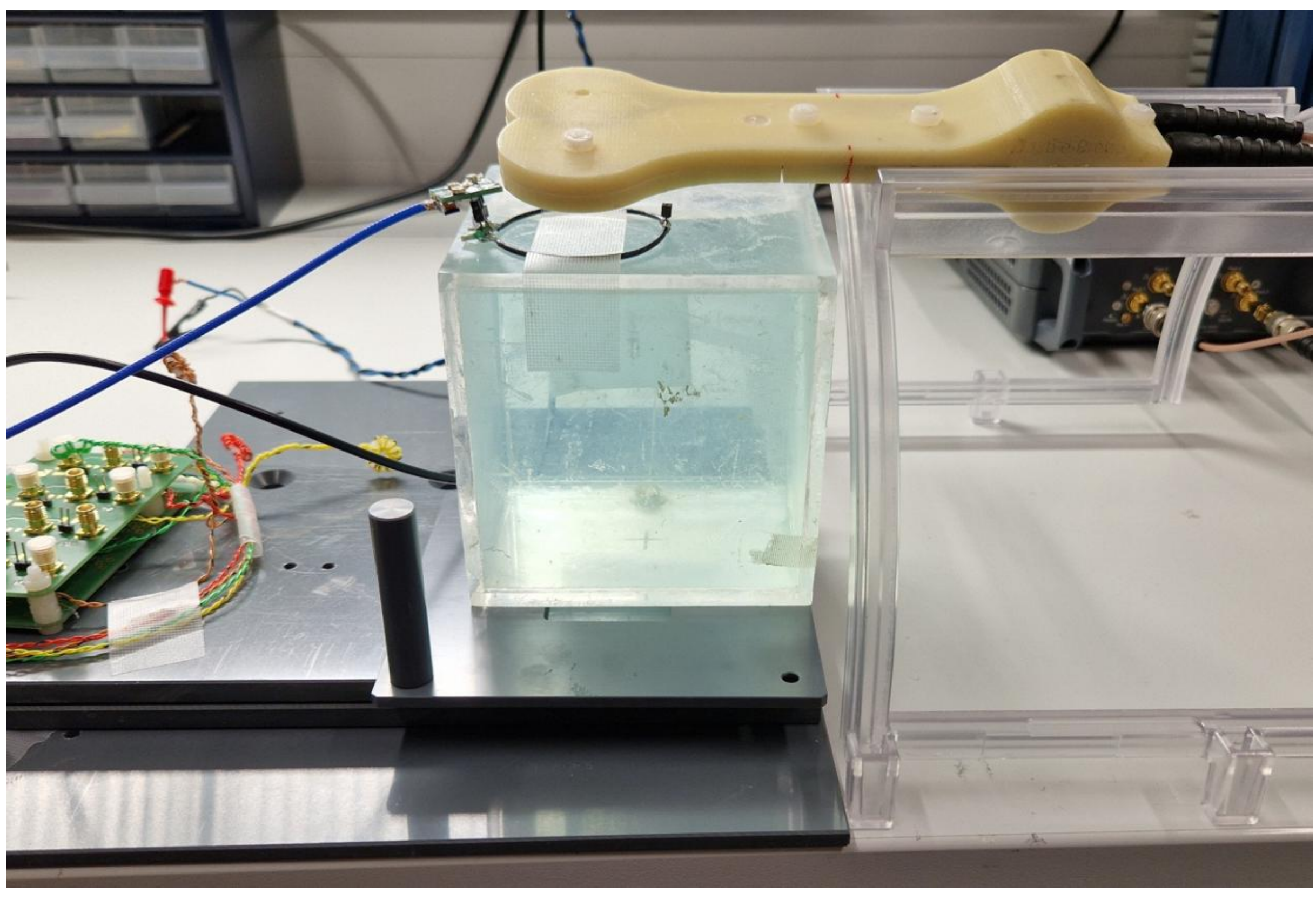

Supporting Information Figure S7: Photo of S21 measurement setup. The investigated coil is fixed on a phantom in the center of the image. Above the coil, a double-loop-probe is fixed at a distance of around 3 cm and connected to a VNA. The port of the coil is connected to a Tee-bias circuit to deliver a galvanic detuning signal.

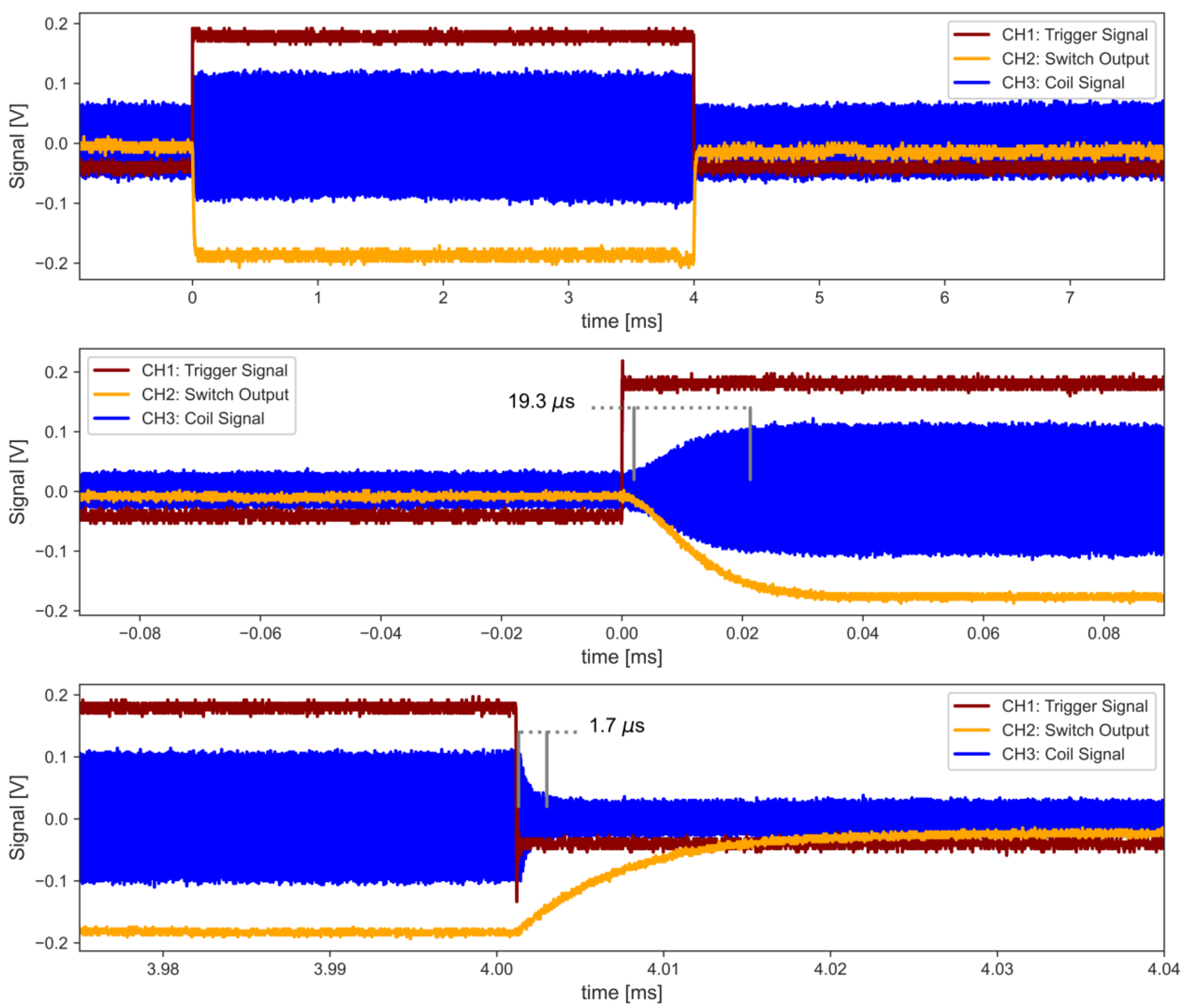

Supporting Information Figure S8: Oscilloscope measurements of PD+PIN tuning and detuning speed with an optical switch (NanoSpeed 1X2, Agiltron, Woburn, MA, USA). An RF signal is coupled from a sniffer loop to a PD+PIN coil, which is connected to Channel 3 of the oscilloscope. A trigger signal is sent to the switch, the output of the switch is split between a simple PD as detector (connected to Channel 2) and the PD+PIN detuning network. At the top is a full cycle of switching between tuned and detuned state. In the middle is a detailed view of the tuning process, which takes about 19.3 µs. At the bottom is a detailed view of detuning, which takes about 1.7 µs and is faster than the switch-off of the direct switch output.